\pdfoutput=1 
\documentclass[aps,twocolumn,preprintnumbers,letterpaper,amsmath,amssymb,10pt]{revtex4-1}
\usepackage{graphicx}
\usepackage[utf8]{inputenc}
\usepackage{scrextend}
\usepackage{manfnt,pifont}
\usepackage{braket}
\usepackage{todonotes}
\usepackage{bbold}
\usepackage{cancel}
\usepackage{pifont}
\usepackage{footmisc}
\usepackage{subfigure}
\usepackage{booktabs}
\usepackage{caption}
\usepackage{graphicx}
\usepackage{amsmath,latexsym}
\usepackage{amsthm}
\usepackage{amssymb}
\usepackage{units}
\usepackage{cases}
\usepackage{xcolor}
\usepackage{fancyhdr}
\usepackage{booktabs}
\usepackage{pgfplots}
\pgfplotsset{compat=newest}
\usepackage{sidecap}
\usepackage{tikz}
\usepackage{tikz-3dplot}
\usepackage{colortbl}
\usepackage{eufrak}
\usepackage{tabularx} 
\usepackage{mathtools}
\usepackage{url}
\usepackage{parskip}
\usepackage{braket}
\usepackage{multirow}
\usepackage{setspace}
\usepackage{listings}
\usetikzlibrary{arrows,shapes,trees,positioning}
\usetikzlibrary{plotmarks}
\usepackage{hyperref}
\usepackage{todonotes}
\newcommand{\zb}{\bar{z}}
\newcommand{\Hpl}[2]{H_{#1}(#2)}

\usepackage{url}

\begin{document}

\preprint{UUITP-3/20}

\title{Towards all-loop supergravity amplitudes on $AdS_5 \times S^5$\\}

\author{Agnese Bissi$^{a}$}
\author{Giulia Fardelli$^{a}$}
\author {Alessandro Georgoudis$^{a,b,c}$}
\affiliation{\it $^{a}$Department of Physics and Astronomy,
	Uppsala University,\\
	Box 516,
	SE-751 20 Uppsala,
	Sweden\\
	\it $^{b}$ Laboratoire de physique de l'Ecole normale sup\'erieure, ENS, Universit\'e PSL, CNRS, Sorbonne Universit\'e, Universit\'e Paris-Diderot, Sorbonne Paris Cit\'e, 24 rue Lhomond, 75005 Paris, France \\
	\it $^{c}$ Institut de Physique Th\'eorique, CEA, CNRS, 
Universit\'e Paris-Saclay, F-91191 Gif-sur-Yvette cedex, France}

\date{\today}

\begin{abstract}
\noindent
We study the four-point function of the superconformal primary of the stress-tensor multiplet in four-dimensional $\mathcal{N}=4$ super Yang Mills theory, at strong coupling and in a large-$N$ expansion. This observable is holographically dual to a four-graviton amplitude in type IIB supergravity on $AdS_5 \times S^5$. We construct the maximal transcendental weight piece of the correlator at order $N^{-6}$ and compare it with the flat space limit of the corresponding two-loop amplitude. This allows us to conjecture structures of the correlator/amplitude which should be present at any loop order.

\end{abstract}



\maketitle

\section{Introduction} 
Since the advent of the AdS/CFT correspondence, the mapping between correlation functions of local gauge-invariant operators and scattering amplitudes has been  in the spotlight. In this paper we address the study of  the four-point function of protected operators of dimension two in four-dimensional $\mathcal{N}=4$ super Yang Mills (SYM) with $SU(N)$ gauge group,  at strong 't Hooft coupling $\lambda=g^2 N$ and as an expansion in inverse powers of $N$. This quantity is holographically related to loop corrections of four-point graviton scattering amplitudes in the supergravity approximation in an $AdS_5 \times S^5$ background. Recently, there has been tremendous progress in understanding how to bootstrap such correlators at order $N^{-4}$ by gluing $N^{-2}$ correlators, using the techniques of the analytic conformal bootstrap \cite{Aharony:2016dwx, Alday:2016njk} and the inversion formula \cite{Caron-Huot:2017vep}. These methods are very reminiscent of  unitarity cuts in amplitudes \cite{Meltzer:2019nbs} and they allowed computing the correlator to order $N^{-4}$ completely \cite{Alday:2017xua, Aprile:2017bgs,Alday:2017vkk, Alday:2019nin}, including also stringy corrections \cite{Alday:2018pdi, Binder:2019jwn, Chester:2019pvm}.  At strong coupling and at leading orders (up to $N^{-2}$) the operators which acquire an anomalous dimension and appear in the operator product expansion (OPE) are double trace operators. These operators are generically degenerate, and the associated mixing problem has been resolved up to order $N^{-2}$ in Refs.~\cite{Alday:2017xua, Aprile:2017bgs, unmixing, DTspectrum}. In this paper we try to understand how much of the four-point function is fixed at a given order $N^{-2\kappa}$ given this information,  namely once we know the OPE data at order $N^{0}$ and $N^{-2}$.  We express, as usual, the four-point correlator  as a function of the cross ratios $U$ and $V$ and focusing on the case $\kappa=3$, we compute the function multiplying the leading logarithmic singularity in $U$. 
In the same spirit as Ref. \cite{Alday:2017vkk} and in order to see how much of the dynamical information we can recover only with this term, we take the flat-space limit and we compare it with the  two-loop four-point supergravity amplitude in ten-dimensional flat space. Quite surprisingly, we notice that the structure of the functions multiplying the highest transcendental pieces \cite{Trans} in the amplitude is the same as the one obtained from the conformal field theory (CFT) computation. More precisely, we find a relation between the CFT flat-space limit and iterated $s$-channel discontinuities of the amplitude.  We conjecture that this fact persists at any loop order and that we can predict certain analytic properties of ladder diagrams, using uniquely the constraints from leading -and subleading-order OPE data. In particular we conjecture that the same identification holds for the full $AdS_5 \times S^5$ space.\\
\section{Four-point function}  
The superconformal primary of the stress-tensor multiplet $\mathcal{O}_{2}$ is a scalar operator of protected dimension two, transforming under the ${\bf{20'}}$ representation of the $SU(4)_R$ R symmetry. The four-point function of $\mathcal{O}_{2}$ has the schematic form
\begin{align} \label{fourpointfn}
\braket{\mathcal{O}_2(x_1)\mathcal{O}_2(x_2)\mathcal{O}_2(x_3)\mathcal{O}_2(x_4)}=\frac{\mathcal{G}(U,V)}{x_{12}^4 x_{34}^4}
\end{align}
where $U,V$ are the cross ratios defined as
\begin{align*}
U=\frac{x_{12}^2 x_{34}^2}{x_{13}^2 x_{24}^2}=z \zb , \qquad\qquad V=\frac{x_{14}^2 x_{23}^2}{x_{13}^2 x_{24}^2}=(1-z)(1-\zb)
\end{align*}
and we disregard $SU(4)_R$ indices for simplicity. 
By using superconformal Ward identities and enforcing unitarity \cite{Nirschl:2004pa,Dolan:2004mu,Beem:2016wfs}, it is possible to disentangle the contribution to the OPE of protected and nonprotected operators, and this allows writing the four-point function as 
\begin{align}
\mathcal{G}(z,\zb)&=\mathcal{G}^{short}(z,\zb)+\mathcal{H}(z,\zb)  \, ,
\end{align}
where $\mathcal{G}^{short}(z,\zb)$ is a known and computable function which repacks the contribution of protected operators, while $\mathcal{H}(z,\zb)$ is a coupling-dependent function and contains information about non protected operators belonging to long multiplets. 
The function $\mathcal{H}(z,\zb)$ admits a decomposition in superconformal blocks $(z \zb)^{\tau/2} g_{\tau+4,\ell}(z,\zb)$ \cite{Dolan:2004iy}
\begin{align} \label{superblocks}
\mathcal{H}(z,\zb)&=\sum_{\tau, \ell} a_{\tau,\ell}(z \zb)^{\tau/2} g_{\tau+4,\ell}(z,\zb) \, ,
\end{align}
where $\tau$ and $\ell$ are respectively the twist (dimension minus spin) and the spin of the intermediate operators which are long superconformal primaries transforming under the singlet of $SU(4)_R$, and $a_{\tau,\ell}$ is the square of the OPE coefficients. \\
The four-point correlator can be expanded around large central charge, $c=\frac{N^2-1}{4}$, as
\begin{multline}\label{Hexp}
\mathcal{H}(z,\zb)=\mathcal{H}^{(0)}(z,\zb)+ c^{-1}\mathcal{H}^{(1)}(z,\zb)\\+c^{-2}\mathcal{H}^{(2)}(z,\zb)+c^{-3} \mathcal{H}^{(3)}(z,\zb)+\dots 
\end{multline}
In the strong-coupling regime we are considering, where the 't Hooft coupling constant $\lambda$ is taken to infinity,  the only single trace operators appearing in the OPE and with finite dimension $p$ are the protected operators $\mathcal{O}_p$ \cite{Alday:2017xua, Aprile:2017bgs}. 
In addition, the set of long operators exchanged to order $c^{-1}$ is only made of double trace operators. They are represented schematically as $[\mathcal{O}_p\, \mathcal{O}_p]_{n,\ell}=(\mathcal{O}_p \Box^{n}\partial_{\mu_1}\dots\partial_{\mu_\ell} \mathcal{O}_p -\text{traces})$ with $p=2, 3, \dots$, and they have classical scaling dimension $\Delta= 2p+2n+\ell$. From this definition it is clear that several operators,  transforming in the same $SU(4)_R$ representation, can  have  the same twist and spin and this fact leads to mixing among $[\mathcal{O}_2\mathcal{O}_2 ]_{n,\ell}, [\mathcal{O}_3\mathcal{O}_3 ]_{n-1,\ell}, \dots, \, [\mathcal{O}_{n+2}\mathcal{O}_{n+2} ]_{0,\ell}$.  The large-$c$ expansion of their OPE data is indeed given by
\begin{align} \nonumber
\tau_{n, \ell}&=4+2n +\frac{1}{c} \gamma_{n,\ell}^{(1)}+\frac{1}{c^2} \gamma_{n,\ell}^{(2)}+\frac{1}{c^3} \gamma_{n,\ell}^{(3)}+ \dots\\  \label{largeNexp}
a_{n,\ell}&=a_{n,\ell}^{(0)}+\frac{1}{c}a_{n,\ell}^{(1)}+\frac{1}{c^2}a_{n,\ell}^{(2)}+\frac{1}{c^3}a_{n,\ell}^{(3)}+\dots
\end{align}
Quite remarkably, the mixing between the aforementioned  double trace operators has been partially solved, up to order $c^{-1}$, so that we eventually know all the $a_{I,n, \ell}^{(0)}$ and $\gamma_{I,n, \ell}^{(1)}$ corresponding to each $I$th eigenstate of the Hamiltonian, with $I=1, \dots, n+1$.
The aim of this paper is to understand how much we can reconstruct of the four-point function at any order $c^{-\kappa}$ with this information by contrasting this piece of the answer with the corresponding flat-space amplitude. \\
\subsection{Method} 
Let us review the method that we are going to use. We are expanding the dynamical part of the four-point function $\mathcal {H}(z, \zb)$ in superconformal blocks as in Eq.~\eqref{superblocks}. Both $a_{n,\ell}$ and $\tau_{n, \ell}$ are meant to be expanded around large $c$, so we plug into the OPE decomposition the expressions Eq.~\eqref{largeNexp}. At any arbitrary order $c^{-\kappa}$, there will be a term,  coming from the $(z \zb)^{\tau/2}$ expansion, of the following form:
\begin{align} \label{higherlog} \nonumber
&\log^{\kappa}(z \zb) \sum_{n, \ell} \sum_{I=1}^{n+1}\frac{a^{(0)}_{I,n,\ell} \left(\gamma^{(1)}_{I,n,\ell}\right)^{\kappa}}{2^{\kappa}\kappa!} (z \zb)^{n+2} g_{8+2n,\ell}(z,\zb)=\\
& \qquad \qquad \log^{\kappa}(z \zb) \frac{(z \zb)^2}{(z-\zb)^{\alpha}}f(z,\zb) \, ,
\end{align}
where $f(z, \zb)$ contains functions of maximal transcendental weight $\kappa$ with polynomial coefficients.  The power  $\alpha$ depends on the order of expansion
\begin{align}
\label{alpha}
\alpha=3\kappa+5(\kappa-1)+4
\end{align}
and is determined by the large-$n$ behaviour of the sum.  Naively one would have expected only the  $3\kappa$ term, which would have reflected  the fact that we are considering $\gamma^{(1)}_{n,\ell}$ to the power $\kappa$ and $\braket{\gamma^{(1)}_{n,\ell}} \to n^3$ as $n \to \infty$.  However this counting is modified by the existence of mixing, which arises from the presence of an R symmetry and reflects the degrees of freedom of the internal manifold. In this way $\alpha$ encloses a contribution from the $AdS_5$ loops and another one coming from the $S^5$.  Finally notice that  the expression in Eq.~\eqref{higherlog} represents the leading logarithmic term of the correlator and it is fully fixed by the  leading and the first subleading OPE data, extracted respectively at order $c^0$ and $c^{-1}$.\\
In general it is possible to use the information obtained at a certain order in the perturbative expansion to get powerful constraints at the next order.  In particular,  thanks to  the Lorentzian inversion formula \cite{Caron-Huot:2017vep},  it is possible to  reconstruct  the full correlation function at any loop order $\kappa$ from its  double discontinuity (dDisc). This can be computed using crossing symmetry and depends on the  information from all the OPE data up to order $\kappa-1$, or equivalently all the functions of $U$ and $V$ in front of  $\log^{\kappa}(z \zb) ,\log^{\kappa-1}(z \zb) \dots \log^{2}(z \zb)$. This method has been  successful at one loop  ($\kappa=2$) \cite{Alday:2017xua, Aprile:2017bgs}, where dDisc does depend only on the $\kappa=0$ and $\kappa=1$ OPE data of double trace operators, for which the mixing has been completely solved.  However already at two loops, it is not possible to fully reconstruct the four-point function due to two obstacles. The first one comes from the appearance of multitrace operators in the OPE, and consequently in dDisc, already at order $c^{-2}$; the second is the presence of mixing which needs to be solved order by order in inverse powers of $c$.  Even if we will not be able to get the full correlator, the knowledge of $a^{(0)}_{I, n, \ell}$
and $\gamma^{(1)}_{I, n, \ell}$ is still enough to completely fix the terms in Eq.~\eqref{higherlog}
for any $\kappa$. In particular  in Ref.~\cite{simon} it has been shown that by rephrasing the problem in terms of ten-dimensional blocks and by acting with a differential operator on them, it is possible to derive a closed form for  this quantity at any order. 
In this paper we will mainly focus on the case  $\kappa=3$ and towards the end we will try to  draw some general conclusions valid for any $\kappa$. \\ 
At two loops the leading logarithmic term is given by
\begin{align} \label{fullres} 
\nonumber
&\mathcal{H}^{(3)}_{\log^3 z \zb}=\frac{(z \zb)^2}{(z-\zb)^{23}} [(R_1 \Hpl{001}{z}+ R_2 \Hpl{1 01}{z}+ R_3 \Hpl{011}{z}\\ & +R_4 \Hpl{01}{z} +R_5 \Hpl{11}{z}+R_6 \Hpl{1}{z}- (z \leftrightarrow \zb ))+R_7] \, ,
\end{align}
where $R_i \equiv R_i(z, \zb)$ are polynomials of degree 30 in $z, \zb$ and $\Hpl{a}{b}$ are harmonic polylogarithms (see  Supplemental Material~\cite{SM} for their definition). Notice that this is only a part of the dDisc, since at this order the correlation function  also contains  a term proportional to $\log^2(z \zb)$, which does contribute to dDisc but that we cannot reconstruct for the reasons listed before.
\\
\subsection{Flat Space}  
To further study the expression in Eq.~\eqref{fullres}, we want to make use of the relation, introduced in  Ref.~\cite{Alday:2017vkk}, between the flat-space limit of the four-point function and the four-graviton scattering amplitude in Minkowski spacetime. In particular we want to understand how much of the full dynamical information we can infer from the knowledge of the highest transcendental weight piece in both sides. 
The key relation between CFT and gravity is given by
\begin{align}
\label{eq:CFTGravityflat}
\lim_{n\to\infty}\frac{\braket{a e^{- i \pi \gamma}}_{n,\ell}}{\braket{a^{(0)}}_{n,\ell}}=b_\ell(s), \qquad L \sqrt{s}= 2 n
\end{align}
where $b_\ell(s)$ are the coefficients of the partial-wave expansion of the gravity amplitude and $L$ represents the AdS radius.  The expression on the lhs  depends on the OPE data in Eq.~\eqref{largeNexp} and can be determined by computing the double discontinuity and then by taking its flat-space limit.\\
The double discontinuity is defined as the difference between the Euclidean correlator and its two possible analytic continuations around $\zb=1$
\begin{align*}
\text{dDisc} \, \mathcal{H}(z, \zb) \equiv \mathcal{H}(z, \zb) -\frac{1}{2}\left( \mathcal{H}^{\circlearrowleft}(z,\zb)+\mathcal{H}^{\circlearrowright}(z, \zb) \right) \, .
\end{align*}
It is useful to apply crossing symmetry and pass to the $t$ channel, where dDisc acts trivially. In our setting, only two terms contribute:
\begin{align} \label{ddiscnv}
\text{dDisc} [\log^2(1-\zb)]&= 4 \pi^2 \\ \label{ddiscnv2}
\text{dDisc} [\log^3(1-\zb)]&= 12 \pi^2 \log(1-\zb) \, .
\end{align}
Notice that with the knowledge of Eq.~\eqref{fullres} we only have access to the part of the double discontinuity coming from Eq.~\eqref{ddiscnv2}.\\
The average $\braket{ae^{-i \pi \gamma}}$ as a function of dDisc  admits a large-$c$ expansion~\cite{SM}, which in the large-$n$ limit takes the form \cite{Alday:2017vkk}
\begin{align} \label{agamma}
 &\nonumber \frac{\braket{a e^{-i \pi \gamma}}_{n,\ell}}{ \braket{a^{(0)}}_{n,\ell}} \xrightarrow{n\gg1}1+ \frac{i \pi n^{3}c^{-1}}{2(\ell+1)}+\frac{c^{-\kappa}}{n^2 (\ell+1)} \int_C \frac{dx}{2 \pi i}e^{-2n x}\\&  \times \int_0^1\frac{d\zb}{\zb^2}\left(\frac{1-\sqrt{1-\zb}}{1+\sqrt{1-\zb}}\right)^{\ell+1}\frac{\text{dDisc}\left[z \zb (\zb-z) \mathcal{H}^{(\kappa)}( z^\circlearrowright,\zb\right]}{4 \pi^2}
\end{align}
where $z^{\circlearrowright}$ stands for  $z \rightarrow z\, e^{-2 \pi i }$ and a sum over all $\kappa \geq 2$ is understood. The integral over $x$, where the contour $C$ encircles the origin clockwise, originates from introducing the flat-space limit $z=\zb+2x \zb \sqrt{1-\zb}$ with $x\to0$.  The powers of $n$ (and consequently of $L\sqrt{s}$ according to Eq.~ \eqref{eq:CFTGravityflat}) produced by this integration are determined by the divergences in $\mathcal{H}^{(\kappa)}$  as $z\to \zb$ and as shown in Eq. ~\eqref{higherlog} this singular behaviour is completely controlled by the power $\alpha$ defined in Eq. ~\eqref{alpha}.  For $\kappa=3$ and restricting to Eq.~\eqref{ddiscnv2}, we get
\begin{equation}\label{dDisctot}
\begin{aligned}
&\frac{\text{dDisc}\left[ z \zb (\zb-z)\mathcal{H}^{(3)} (z^{\circlearrowright}, \zb) \right]}{4 \pi ^2}\\ & \qquad \qquad \qquad \qquad \rightarrow 2 \pi i \frac{\Gamma(22)}{(2x)^{22}} \frac{\Hpl{1}{\zb}}{2^{10}(15)^3}r_3(\zb)\, ,
\end{aligned}
\end{equation}
where 
\begin{align} \nonumber
&r_3(\zb)=\frac{60(1-\zb)^6}{ \zb^6 } \Big\lbrace  \frac{p_4^{(a)}-p_4^{(b)}}{2} \Hpl{1}{\zb}^2- p_4^{(b)} \Hpl{0}{\zb} \Hpl{1}{\zb} \\& \nonumber  \,\,+ \left(p_4^{(a)}+p_4^{(b)}\right) \Hpl{01}{\zb}+ \left(\frac{p_3^{(a)}-p_3^{(b)}}{60}+ i \pi
    p_4^{(b)}\right) \Hpl{1}{\zb} \\  & \label{eq:CFTg3} \,\, -\frac{p_3^{(b)}}{60} \Hpl{0}{\zb}+\frac{i \pi  p_3^{(b)}}{60}+\frac{\pi ^2 p_4^{(a)}}{6}+q_2(\zb)\Big\rbrace \, .
\end{align}
The explicit expressions for the polynomials $p_i, \, q_2$ can be found  in the Supplemental Material~\cite{SM}.  Let us conclude by mentioning that it is possible to construct the polynomials analogous to $p_4$ and $p_3$  at any loop order,  we leave this result to Ref. ~\cite{Bissi:2020woe}. \\
\section{Amplitude} 
As we have seen in the flat-space limit of AdS$_5 \times$S$^5$ we obtain ten-dimensional supergravity.  The four-point function we consider in Eq.~\eqref{fourpointfn} is dual to the ten-dimensional amplitude of four gravitons. Up to two loops, this schematically takes  the form \cite{Bern:1998ug}
\begin{widetext}
\begin{align}
\label{10dAmpl} \nonumber
\mathcal{A}_{10}^{sugra}&=\hat{K} \left\lbrace\frac{8 \pi G_N}{s t u}+\left(8 \pi G_N\right)^2\left(I_{box}(s,t)+I_{box}(t, s)+I_{box}(t,u) \right) + \left(8 \pi G_N\right)^3\left(s^2\left(I_{db}^{pl}(s,t)+I_{db}^{np}(s,t)+t\leftrightarrow u\right)\right.\right.  \\& \left. \left.+t^2\left(I_{db}^{pl}(t,s)+I_{db}^{np}(t,s)+s\leftrightarrow u\right)
+u^2\left(I_{db}^{pl}(u,s)+I_{db}^{np}(u,s)+s\leftrightarrow t\right) \right)+ \mathcal{O}(G_N^4 ) \right\rbrace\\
& \equiv (\pi L)^5 s^4\Big\lbrace \frac{L^3 f_1(x)}{s^3 c}+\frac{L^{11} s f_2(x)}{c^2}+\frac{L^{19} s^5 f_3(x)}{c^3} \Big\rbrace +  \mathcal{O}(c^{-4} ) 
\end{align}
\end{widetext}
where $I_{box}, \, I_{db}^{pl}$ and $I_{db}^{np}$ represent respectively the single and planar/nonplanar double box. In the third line we have used the identification $8\pi G_N=\pi^5L^8c^{-1}$. $\hat{K}$ is a dimension-eight kinematic factor, depending on graviton polarization and fixed to be $s^4$ in this case. 

In order to evaluate  $I_{db}^{pl}$ \cite{Bern:1998ug, Smirnov:1999gc,Smirnov:1999wz} we have first computed the integral in four dimensions using the differential equation method proposed in Refs.~\cite{Kotikov:1990kg, Remiddi:1997ny, Henn:2013pwa},  and then we have uplifted the result to $10-2 \epsilon$ via the dimensional recurrence relation \cite{Tarasov:1996br,Lee:2009dh}. The final expression can be found in the Supplemental Material~\cite{SM}.
It is important to remark that  $I_{db}^{np}$ has lower maximal transcendental weight and as so it will not contribute to our computation \cite{footnote2}.

The quantity that enters Eq.~\eqref{eq:CFTGravityflat}, through its partial-wave expansion,  it is not directly $\mathcal{A}_{10}^{sugra}$ but $\mathcal{A}_5$,  i.e. the first one divided by the volume of $S^5$. The partial-wave expansion then reads
\begin{align}
i \mathcal{A}_5(s,t)=\frac{128\pi}{\sqrt{s}} \sum_\ell (\ell+1)^2 b_\ell(s)P_\ell(\cos \theta)
\end{align}
where $P_l$ are Legendre polynomials. The $b_\ell$ admits a form similar to Eq.~\eqref{agamma}
\begin{align}\label{bl_amp}
\nonumber
&b_\ell(s)=1+ \left(\frac{L\sqrt{s}}{2}\right)^3\frac{i \pi c^{-1}}{2(\ell+1)}+\int_0^1\frac{d\zb}{\zb^2}\left(\frac{1-\sqrt{1-\zb}}{1+\sqrt{1-\zb}}\right)^{\ell+1}\\ &  \quad \times \frac{2 \pi i}{\ell+1} \left(\text{disc}_t+(-1)^\ell\text{disc}_u \right) \frac{\sqrt{s}\mathcal{A}_5(c,s, \cos\theta)}{64\pi^2}
\end{align} 
where we have introduced a dispersion relation for the amplitude \cite{Caron-Huot:2017vep,Alday:2017vkk}.  The similarity between this expression and Eq.~\eqref{agamma} motivates the identification in Eq.~\eqref{eq:CFTGravityflat} and allows comparing directly dDisc and discontinuities. Notice that in our case $\text{disc}_t \mathcal{A}(s,t,u)=\text{disc}_u \mathcal{A}(s,t,u)=-\text{disc}_s \mathcal{A}(s,t,u)$ so we can further simplify  Eq.~\eqref{bl_amp}.\\
At two loops, we see from Eq.~\eqref{10dAmpl} that the amplitude is constructed from the sum over the three different channels, $s,t$ and $u$. To compare with our results for  the CFT dDisc in Eq.~\eqref{dDisctot}, it is enough to study the $t$-channel contribution to the discontinuity in $t$, which we can compute as
\begin{equation}\label{disct_amp}
 \text{disc}_t \mathcal{A}^t =\hat{K} \left(8 \pi G_N \right)^3 t^2 \text{disc}_t \left( I_{db}^{pl}(t,s)+s\leftrightarrow u \right).
\end{equation} 
To have a better interpretation of this discontinuity, it can be  useful  to construct $\text{disc}_y I_{db}^{pl}$   diagrammatically. This is done by summing over all possible cuts in which $y$ is the generic momentum flowing. A given cut diagram, in turn, can be constructed from the integral representation of the diagram by putting  the cut propagators on shell \cite{Cutkosky:1960sp}. As in our case the discontinuities in the different channels are related we can focus on  $\text{disc}_s$ :
\begin{figure}[h!]
\includegraphics[width=0.32\textwidth]{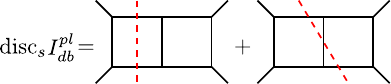}
\end{figure}\\
With this approach, it is evident that we have the contribution of two different types of cuts: a double one ($c_1$) and a triple one ($c_2$).
A way to  extract these is to construct and solve a system of  differential equations on $c_1$ and $c_2$ \cite{Bosma:2017hrk}. {To solve it, some input boundary conditions are needed and these can be fixed in such a way that we get \cite{Bissi:2020woe}}
\begin{align}\label{amp2}
&\text{disc} _s I_{db}^{pl}=2 \pi i \left(I_{db}^{pl}\big|_{c_1}+I_{db}^{pl}\big|_{c_2}\right) \quad \text{with}\\
\nonumber& I_{db}^{pl}\big|_{c_1}=\frac{s^4 \left( 60 p_4(s,t)\Hpl{-100}{x}+p_3(s,t)\Hpl{00}{x} \cdots \right)}{2^5(15)^3t^2}\, ,\\
\nonumber &I_{db}^{pl}\big|_{c_2}=\frac{s^4}{2^5(15)^3t^2}( 60 p_4(s,t)\left(-\Hpl{-100}{x}+\Hpl{-1-10}{x}\right) \\ & \nonumber + p_3(s,t)\left(-\Hpl{00}{x}+\Hpl{-10}{x}\right) \cdots)\,. 
\end{align}
where $x=\frac{t}{s}$. This pictorial representation of the discontinuity suggests that the two cuts should have different CFT counterparts.  In particular, in the same light of Ref.~\cite{Meltzer:2019nbs}, we interpret  $c_1$  as an exchange of only double trace operators, while $c_2$ is an exchange of triple trace ones. Given this insight and considering that Eq.~\eqref{fullres} only includes double trace contributions, we construct a similar object in the amplitude. Graphically, this corresponds to
\begin{figure}[h!]
\includegraphics[width=0.13\textwidth]{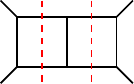}
\end{figure}\\
This double $s$-channel cut can be computed by means of differential equations and it reads:
\begin{align}
\label{dc}
I_{db}^{pl}\big|_{dc}=\frac{\chi s^4 \left(60 p_4(s,t) \Hpl{-10}{x}+p_3(s,t)\Hpl{0}{x}+\cdots \right)}{2^6(15)^3 t^2}
\end{align}
where $\chi$ is a normalization constant we can not fully  fix.\\
\section{Comparison} 
With the main ingredients for the $\kappa=3$ computation in hand, we can now compare the partial result from the CFT side with the flat-space amplitude.  First of all we need to map the results of Eqs.~\eqref{amp2} and \eqref{dc} to the CFT notation by using the following redefinition $x=\frac{1-\zb}{\zb}$ \cite{Alday:2018pdi}. 
Plugging the CFT results~\cite{SM} into  Eq.~\eqref{eq:CFTGravityflat}, we get
\begin{widetext}
\begin{align}
&\frac{\braket{a e^{-i \pi \gamma}}_{n,\ell}}{ \braket{a^{(0)}}_{n,\ell}}\to 1+ \frac{i \pi n^{3}}{2c(\ell+1)}+\frac{i \pi}{\ell+1} \int_0^1\frac{d\zb}{\zb^2}\left(\frac{1-\sqrt{1-\zb}}{1+\sqrt{1-\zb}}\right)^{\ell+1} \left(\frac{n^{11}}{c^2}g_2(\zb)+\frac{n^{19}}{c^3}\left(\frac{1-\zb}{\zb}\right)^6(p_4^{(a)}-p_4^{(b)})\Hpl{1}{\zb}^3 +...\right)\, , \nonumber \\ 	\label{bl}
\end{align} 
\end{widetext}
where $g_2(\zb)$ is defined as \cite{Alday:2017vkk}
\begin{equation}
\begin{aligned}
&g_2(\zb)=\frac{(1-\zb)^2}{960 \zb^4}\left(2(1-\zb^5)\Hpl{1}{\zb}- 2\zb^5 \Hpl{0}{\zb}\right. \\  &\qquad \qquad \qquad \quad \left.  +2i \pi \zb^5+ \zb \left(2 \zb^3+\zb^2-\zb-2\right)\right) \, .
\end{aligned}
\end{equation}
At order $c^{-3}$ we have reported only the highest $\log$ piece of the dDisc of the correlator, which  appears in the amplitude discontinuity as well. 
It is possible to show that, given Eq.~\eqref{amp2},  this term can only come from the $c_1$ type of cut.  If we restrict to just the  $c_1$ contribution, constructed from Eqs.~\eqref{disct_amp} and \eqref{amp2}, the amplitude  discontinuity can be written  as
 \begin{align} \nonumber
&\mathcal{A}^{t}\big|_{c_1}  \propto \frac{(1-\zb)^6}{\zb^6} \Big\lbrace \frac{p_4^{(a)}-p_4^{(b)}}{2} \Hpl{1}{\zb}^3- \frac{3}{2}p_4^{(b)}\Hpl{0}{\zb} \Hpl{1}{\zb}^2 \\&  + (\frac{p_3^{(a)}-p_3^{(b)}}{20}+3 i \pi  p_4^{(b)}) \Hpl{1}{\zb}^2 -\frac{p_3^{(b)}}{20} \Hpl{1}{\zb} \Hpl{0}{\zb} \Big\rbrace \,.
\end{align}
We notice that it exactly reproduces the functional form of  Eq.~\eqref{eq:CFTg3}.  These results further support our interpretation of the cuts in terms of multitrace operator exchange.  The mismatch in the numerical prefactors that we see is due to the fact that ideally $\mathcal{A}^{t}\big|_{c_1}$ should reproduce the full contribution of double trace operators to the dDisc of the correlator and Eq.~\eqref{dDisctot} is only part of it.  The unknown $\log^2 U$ piece of the CFT correlator should thus have a very specific form in order to solve these discrepancies.\\
Given this insight and the fact that the term $\mathcal{H}^{(3)}$ in Eq.~\eqref{fullres} is constructed from $(\gamma^{(1)})^3$, we argue that to extract the same contribution in the amplitude one has to consider  the double cut $\mathcal{A}^{t}\big|_{dc}$ in Eq.~\eqref{dc}.  And indeed we find 
\begin{align}
\mathcal{A}^{t}\big|_{dc}=32 s^5 r_3 \, ,
\end{align}
so that the amplitude and the CFT computations match perfectly once we have factored out a logarithm as in Eq.~\eqref{dDisctot}.\\

Given these results we can now make some general considerations:
\begin{itemize}
\item We conjecture that at all loops the highest $\log U$ contributions can be extracted from the saturated $(\kappa-1)$ $s$-channel cuts
\begin{figure}[h!] \hspace{0.6cm}
\includegraphics[width=0.25\textwidth]{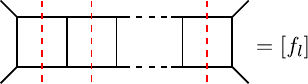}
\end{figure}        \\
and moreover this contribution should show a log factorization as $g_{\kappa}=\Hpl{1}{\zb}^{\kappa-2} \left[f_l \right]$, where $f_l$ is a function of maximum weight $l$ at any $l$ loop order, $l=\kappa-1$. 
\item Since the leading log contribution is obtained from tree-level data we can make some predictions on the powers of $\log^n U\log^k V $ appearing at higher orders.
A heuristic argument can be made by considering two particle cuts and how they, depending on the momentum being cut, will contribute to a $\log$ of $U$ or $V$ times a lower loop diagram.  
 With this in mind we conjecture that the highest log contribution in the correlation function is of the form $\log^{\kappa} U\log^2 V $. Subleading log contributions can be extracted from generalized ladder integrals in a similar fashion, for example for $\kappa=4$ we conjecture the following behaviour:
\begin{figure}[h!] \hspace{0.4cm}
\includegraphics[width=0.3\textwidth]{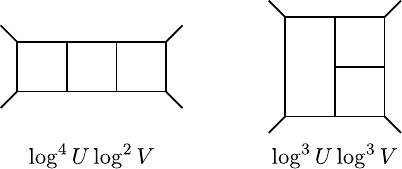}
\end{figure}
\item We strongly believe that the identification of the iterated $s$ cut of the ladder diagram with the piece of the correlator in Eq.~\eqref{higherlog} also persists in curved space and to all loops. 
\end{itemize}
\vspace{1cm}
\begin{acknowledgments}
We thank F. Alday for several discussions and insights, and we thank F. Alday, P. Dey and T. Hansen for comments in the draft. AG also thanks B. Page, S. Abreu, B. Basso, E. Trevisani, V. Goncalves, R. Pereira and H. Paul for useful discussion. 
The work of AB and GF is supported by Knut and Alice Wallenberg Foundation under grant KAW 2016.0129 and by VR grant 2018-04438.
The work of AG is supported by the Knut and Alice Wallenberg Foundation under grant 2015.0083 and by the French National Agency for Research grant ANR-17-CE31-0001-02 and ANR-17-CE31-0001-01.
\end{acknowledgments}

\newpage
\begin{widetext}
\section*{Supplemental materials}
\subsection*{Harmonic Polylogarithms}
The  Harmonic Polylogarithm (HPL) \cite{Remiddi:1999ew} is defined iteratively as: 
\begin{equation*} \label{hpl}
\phi(0,x)=\frac{1}{x}  \quad \phi(1,x)=\frac{1}{1-x} \quad \phi(-1,x)=\frac{1}{1+x} \,,
\end{equation*}
\begin{equation*}
H_{\,a,\cdots}= \int_0^x \text{d}x^\prime H_{\cdots}(x^\prime) \phi(a,x^\prime)\,.
\end{equation*}
From this definition it is straightforward to recover the usual Polylogarithms as Li$_n=H_{\vec{0}_{n-1}1}(x)$.

\subsection*{Expansion of the average} 
The average  $\braket{ae^{-i \pi \gamma}}_{n,\ell}$ in (12) can be expanded for large $c$ in the following way
\begin{align*}
&\braket{ae^{-i \pi \gamma}}_{n,\ell}=\braket{a^{(0)}}+c^{-1}\left( \braket{a^{(1)}}-i \pi\braket{a^{(0)}\gamma^{(1)}} \right)+  c^{-2}\left(\braket{a^{(2)}}-i \pi\braket{a^{(1)}\gamma^{(1)}+a^{(0)}\gamma^{(2)}}-\frac{\pi^2}{2}\braket{a^{(0)}{\gamma^{(1)}}^2}\right)\\  & \quad
 +c^{-3}\big(\braket{a^{(3)}}-i \pi\braket{a^{(2)}\gamma^{(1)}+a^{(1)}\gamma^{(2)}+a^{(0)}\gamma^{(3)}}-\pi^2\braket{\frac{a^{(1)}{\gamma^{(1)}}^2}{2}+a^{(0)}\gamma^{(1)}\gamma^{(2)}}+\frac{i \pi^3}{6}\braket{a^{(0)}{\gamma^{(1)}}^3} \big)+O(c^{-4})
\end{align*}
where the bracket refers to the average over the degeneracy index $I$.\\
\subsection*{Polynomials} 
Here we collect the explicit expression for the polynomials $p_{3,4}$ and $q_2$ appearing in (13).
\begin{align*} \label{p34}
&p_4(a,b)=\frac{21 a^2+12 a b+b^2}{b}\\
&p_3(a,b)=-\frac{1}{(a+b)^3}(1260 a^4+3870 a^3 b+4170 a^2 b^2\\ &\qquad+1785 a b^3+227 b^4)\\
&q_2(\zb)=\frac{(1258 \zb^3-871  \zb^2+871 \zb-1258)}{60\zb}
\end{align*}
In the main text we have defined $p_{3,4}^{(a)}\equiv \zb p_{3,4}(\zb-1,1)$, $p_{3,4}^{(b)}\equiv p_{3,4}\left(\frac{1-\zb}{\zb},1\right)$.
\subsection*{Ten Dimensional Amplitude} 
The full expression for the planar double box $I^{pl}_{db}(s, t)$, depicted below, 
\begin{figure}[h!]
\label{fig:doublebox} 
\centering
\begin{tikzpicture}[scale=0.4]
    \draw [thick] (0,0) -- (6,0)--(6,3)--(0,3)--(0,0);
    \draw [thick] (3,0) -- (3,3);
    \draw [thick] (0,0) -- (-1,-1);
    \draw [thick] (0,3) -- (-1,4);
    \draw [thick] (6,0) -- (7,-1);
    \draw [thick] (6,3) -- (7,4);
    \draw [thick,->] (3,0) -- (4.5,0);
    \draw [thick,->] (3,0) -- (1.5,0);
    \node [below left=0.1 cm] at (-1,-1)  {$p_1$};
    \node [below right=0.1 cm] at (7,-1)   {$p_4$};
    \node [above right=0.1cm] at (7, 4) {$p_3$};
    \node [above left=0.1 cm] at (-1,4)  {$p_2$};
    \node [below=0.1 cm] at (1.5,0)  {$l_1$};
    \node [below=0.1 cm] at (4.5,0)  {$l_2$};
\end{tikzpicture}
\end{figure}
is given by
\begin{align*}
&I_{db}^{pl}(s,t)=\frac{s^2(4s+t)}{7!60\epsilon^2}+\frac{1}{7!^2 10 \epsilon} (22384 s^3+6247 t s^2 +63 t^2 (4 s+t))-\frac{s^4}{2^5(15)^3t^2}(60 p_4(s,t)\Hpl{-1-100}{x}+  \\  \nonumber  & \qquad  p_3(s,t)\Hpl{-100}{x} +p_2 \Hpl{00}{x} +p_1 \Hpl{0}{x} +p_0 \cdots)
\end{align*} 
where we have factored out the scale dependence and we have denoted with $p_{i}(s,t)$ the polynomials multiplying transcendentality $i$ functions.  Notice that the order $\epsilon^{-1}$ does not match exactly the one in \cite{Bern:1998ug} because we have not explicitly subtracted the one loop pole. In the $\epsilon^0$ part we have reported explicitly only the highest transcendental weight functions, as those will be relevant in the main discussion.\\

\end{widetext}

\begin{thebibliography}{33}
\bibitem [{foo()}]{footnote2}%
  \BibitemOpen
  \href@noop {} {}\bibinfo {note} {It will also be clearer later from the
  possible dispersion relation that this has to be the case, since this diagram
  does not allow a double $s$-channel cut.}\BibitemShut {Stop}%
\end{thebibliography}
\end{document}